\documentclass[preprint,showpacs,preprintnumbers,amsmath,amssymb]{revtex4}

\usepackage{graphicx}
\usepackage{dcolumn}
\usepackage{bm}

\begin{document}


\title{A Physical Axiomatic Approach to Schrodinger's Equation}

\author{Rajesh R. Parwani}
 \email{parwani@nus.edu.sg}
\affiliation{Department of Physics and University Scholars Programme\\ National University of Singapore,\\
Kent Ridge, Singapore.}


\vspace{0.3in}

\begin{abstract}
The Schrodinger equation for non-relativistic quantum systems is derived from some classical physics axioms within an ensemble hamiltonian framework. Such an approach enables one to understand the structure of the equation, in particular its linearity, in intuitive terms. Furthermore it allows for a physically motivated and systematic investigation of potential generalisations which are briefly discussed.

\end{abstract}

\maketitle

\section{Motivation}

It is often stated that of the two departures from Newtonian physics at the beginning of the twentieth century, relativity theory has a pleasing physical foundation while quantum theory is grounded more in abstract mathematical structures.  Textbooks either quote the Schrodinger equation with little motivation or obtain it as the description of  state evolution in a particular picture: but the reason for choosing states in a linear vector space in the first place is left unexplained.

Schrodinger's original derivation, involving analogies with wave optics and various limits, is now considered only of heuristic value as he was then still unaware of the interpretation of the wavefunction as a probability amplitude rather than a physical wave. However, with hindsight, we know that Schrodinger's equation may be re-written in more familiar terms through a change of variables; starting from
\begin{equation}
i\hbar \dot{\psi} = \left[ - {\hbar^2 \over 2m}  \partial^2 + V \right] \psi \, , \label{sch1}
\end{equation}
one performs a Madelung transformation \cite{mad} $\psi = \sqrt{p} \ e^{i \sigma / \hbar}$ which decomposes the Schrodinger equation into two real equations,
\begin{eqnarray}
\dot{\sigma} + {1 \over 2m} (\partial_i \sigma)^2 + V +{ {\hbar}^2 \over 8m}  \left( { (\partial_i p)^2  \over p^2 } -{2 \partial_{i}^{2} p \over p}  \right) &=& 0 \, ,  \label{hj3} \\
\dot{p}  + \ { 1\over m} \partial_i \left( p \ \partial_i \sigma \right)  &=& 0 \, ,  \label{cont3} 
\end{eqnarray}
the summation convention being used unless otherwise stated and the overdot referring to a partial time derivative. 
The first equation is a generalisation of the classical Hamilton-Jacobi equation, the term with explicit $\hbar$ dependence, the ``quantum potential", summarising  the peculiar aspects of quantum theory. 
If the quantum potential is ignored then the equations have a simple classical interpretation: It is assumed that one is uncertain about the initial conditions so that probabilistic methods must be used to describe the location of the particle. With $p(x,t)$ denoting the normalised probability density, the second equation of motion above is the continuity equation with $\sigma$ determining the velocity, $v_i$,  through $v_i = (\partial_i \sigma)/m$. What transforms the classical ensemble dynamics into quantum mechanics is the quantum potential, the point of focus in the deBroglie-Bohm picture \cite{BH}.

However the structure of the quantum potential is unusual, making the quantum Hamilton-Jacobi equation (\ref{hj3})  difficult to understand in purely classical terms. Though some studies, such as those of Nelson \cite{Nelson} and others \cite{smolin}, have been made to derive (\ref{hj3})  from some stochastic micro-dynamics, the assumptions either  go beyond familiar classical physics or introduce additional ingredients that raise new puzzles. 

A somewhat different approach has been to start with the plausible classical ensemble Hamilton-Jacobi equation, argue that it is incomplete, and then try to constrain possible extensions by some consistency requirements. Two such recent derivations are in \cite{HR1,P1}.
The assumption of Hall and Reginatto was that the classical equation only described the mean motion of the particles, and that the momentum of the particles have some fluctuations about the mean value. It was postulated that those fluctuations obey an exact uncertainty relation \cite{Hall}, and that the fluctuation term  also obeyed some axioms such as locality and separability. In \cite{P1} on the other hand, the maximum uncertainty (entropy) principle \cite{Jay} was used, as suggested earlier in \cite{Reg,fried}, to constrain the probability distribution $p(x,t)$. The constraint was implemented through a lagrange multiplier and the unique uncertainty measure that accompanies the lagrange multiplier was constructed from physically motivated axioms. The end result in both approaches is that the classical Hamilton-Jacobi equation gets a contribution, the  quantum potential, with Planck's constant making its appearance to balance the dimensions between the old terms and the new.

Once the quantum Hamilton-Jacobi equation (\ref{hj3}) has been obtained, it and the continuity equation (\ref{cont3}) represent two coupled nonlinear differential equations for the variables $p,\sigma$. They can be uncoupled and linearised through the  Madelung transformation resulting in the usual Schrodinger equation and its complex conjugate. The meaning behind the  Madelung transformation was explained in \cite{HR1}: it is a change to a canonically conjugate set of variables that would uncouple the equations.

Thus in both of the approaches \cite{HR1,P1}, only the {\it extension} of the classical equation  was constructed from various axioms: this still gives the impression that something extra and special must be imposed on classical ensemble dynamics to arrive at quantum theory. However, as this paper aims to demonstrate, one can proceed much further.

The primary purpose of this paper is to present a  set of physical axioms that can be used to construct the Schrodinger's equation directly, without assuming the classical ensemble Hamilton-Jacobi equation as the starting point, nor assuming any specific underlying dynamics.  For this a Hamiltonian framework for ensembles, as discussed in \cite{HR1,HR2}, will be used but the axioms will be  refined from those used in \cite{P1}. This approach will achieve three goals: (i) It will show that it is possible to understand the structure of  Schrodinger's equation in standard classical physics terms, (ii) emphasize that one may arrive at
a quantum theory without ``quantising", in one way or another, some classical dynamics (which logically should be the limit of the quantum theory) and (iii) provide an avenue for physically motivated extensions of quantum theory that might be relevant for current studies of space at short distances. 

The axioms are listed and explained in the next section followed by the construction of the ensemble hamiltonian in Sect.(3). The main differences between Ref.\cite{P1} and this paper are discussed in the concluding section together with some comments on potential generalisations of linear quantum theory. Some technical issues concerning Galilean invariance and gauge-inequivalence are discussed in the appendices.

\section{The Axioms}
We wish to describe the dynamics of $N$ particles of which  we do not have sufficient information about the initial conditions, so that statistical methods must be used to locate the particles. Note that even for a single particle, $N=1$, one has an ensemble dynamics. The discussion is being carried out for the general multiparticle case so that the separability axiom can be discussed. 

Let $p(x,t)$ denote the normalised probability density for the $N$ particles, with $x$ summarising all the spatial coordinates. The following discussion will use Cartesian coordinates in $d+1$ dimensions, with configuration space indices $i,j= 1,2,......,dN$. Here $i=1,...d$, refer to the coordinates of the first particle of mass $m_1$, $i=d+1,.....2d$, to those of the second particle of mass $m_2$ and so on. A diagonal and positive definite configuration space metric, $g_{ij} = \delta_{ij} /m_{(i)}$, with the symbol $(i)$ defined as   the smallest integer $\ge i/d$, is assumed as in classical dynamics \cite{Reg}.  That metric not only encodes information about the inertia (mass) of the particles, which can be different, the indices allow contraction with derivatives and enable a useful summary of the spacetime symmetries that will be assumed below.

Let $H$ be the ensemble hamiltonian depending on the normalised probability density $p(x,t)$ and let $S(x,t)$ denote the canonically conjugate variable, 
\begin{equation}
H \equiv \int dx^{Nd} p \ ( h(p,S) + V), \label{ensHam}
\end{equation}
with $V$ some external potential influencing the particles' motion. Hamilton's equations are
\begin{eqnarray}
{\partial p \over \partial t} &=& {\delta H \over \delta S} \; ,\label{ham1} \\
{\partial S \over \partial t} &=& -{\delta H \over \delta p} \; . \label{ham2} 
\end{eqnarray}
Since $p$ is the probability density, the first equation will turn out to be the continuity equation.

The objective is to construct an explicit form for $h(p,S)$ in (\ref{ensHam}) starting from the following reasonable  axioms:

\begin{itemize}
\item $[A0] \;$ {\bf Hamiltonian}: The existence of an ensemble hamiltonian of the form (\ref{ensHam}),  evolution determined by Hamilton's equations,  and a  configuration space metric with the stated properties, may be formalised in this opening axiom. 

\item $[A1] \;$ {\bf Locality}: $h$ should be a function of $p,S,$ and their {\it spatial} derivatives. That is, for example, $h$ should not involve any integrals as that would couple fields at distant points and create problems with causality. Note that disallowing time-derivatives in $h$ ensures that hamilton's equations (\ref{ham1},\ref{ham2}) have time-derivatives only on the left-hand-side.  

\item $[A2] \;$ {\bf Separability}: $h$ should be separable for the case of two independent sub-systems described by probability distributions $p_1$ and $p_2$; $h(p=p_1 p_2) =h(p_1) + h(p_2)$ so that $H$ itself can be written as the sum of two independent terms. The factorisation of $p$ for separable systems will affect how $S$ behaves through Hamilton's equations; it is sufficient for consistency to require $S=S_1 + S_2$.  

\item $[A3] \;$ {\bf Symmetry}: The observed symmetries of Nature should be included in the description of the system. Since we are working in the non-relativistic limit, the equations of motion should be form-invariant under the Galilean group\footnote{The symmetries refer to the case of vanishing external potential, $V=0$.}. The translational part of this symmetry can be used to eliminate any explicit dependence of $h$ on $x_i,t$ while rotational invariance will be used as an explicit constraint in the construction below. However invariance under Galilean boosts can only be imposed on the equations of motion, rather than on $h$, as that involves transformations of the time-derivatives: remarkably though, the equations that are obtained below, after imposing the other conditions, are already invariant under boosts (see Appendix A).

\item $[A4] \;$ {\bf Universality}: It is desirable to construct a hamiltonian that describes universal dynamics, that is, $h$ should be independent of $V$ and any other specific properties of the particles or their number except those encoded in the configuration-space metric. An example of how this universality constraint can be used is as follows: the normalisation of probability, $1 = \int dx^{Nd} p(x,t)$, implies that the dimension of $p(x,t)$ depends on the dimension of the configuration space. Thus, as elaborated below, $H$ can be universal only if $h$ is scale invariant, $h(\lambda p) =h(p)$. This scale invariance will ensure that the resulting equations of motion have a form independent of the number of particles. Hamilton's equations show that the scale-invariance condition does not affect $S$.

It must be emphasized that universality as defined above implies much more than simply the statement that the terms in $h$ be independent of the dimension of configuration space. For example, to compensate for the dimension of $p$ one might just try using $N$ powers of $x$, but since translational invariance (for $V=0$) disallows explicit dependence on $x$, one must resort to derivatives. So consider the candidate $h= p^{-1} (\partial_1 \partial_2......\partial_N \log p) $: its dimension is independent of the dimension of configuration space and yet it is not scale invariant. However this example  clearly does not lead to {\it universal} dynamics because the form of the equations of motion change as the number of particles changes, with higher-derivative terms appearing with increasing $N$. This shows that imposing universality fully, as defined, clearly disallows compensating the dimension of $p$ with factors of $x$ or $\partial/\partial x$. Which means that $H$ can be universal only if $h$ is scale invariant, as stated above.

A beneficial technical outcome from the scale invariance of $h$ is that one need not impose the normalisation condition on $p$ in the hamiltonian but rather postpone it to a latter stage, for example after solving the equations of motion.  

\item $[A5] \;$ {\bf Positivity}: The existence of a stable ground state requires that the hamiltonian $H$  be bounded from below for potentials $V$ that are likewise bounded. For this to be true for generic potentials, and thus lead to universal dynamics, it is sufficient to choose $h$ to be positive definite. 
 
\item $[A6] \;$ {\bf Simplicity}: The fewer arbirary parameters a scientific theory has, the easier it can be falsified experimentally and so lead to suggestions for improvement. Furthermore, if a simple theory is able to explain all the available data then one is more likely to accept it as something ``fundamental" rather than an intermediate phenomenological description. Indeed, a useful working hypothesis in physics has been that fundamental laws should be universal and simple. 

Thus one would like to construct a hamiltonian that has a minimum number of arbitrary constants. For example, the scale invariance of $h$ deduced in $[A4]$ means that $h$ must contain derivatives of $p$ if it is to depend on $p$. Rotational symmetry implies that at least two derivatives would be required. Therefore one constraint is to let $h$ contain not more than two derivatives in any product of terms that appears in it. As each derivative involves an inverse length, this condition obviously restricts the number of new dimensional parameters, beyond the metric,  that can appear in the action. This specific implementation of the simplicity condition, whereby not more than two derivatives appear in any product of terms, will be referred to in brief as ``absence of higher number of derivatives" or ``AHD". 

It is implicit in the foregoing discussion  that one would like to construct and describe some realistic dynamics and so the simplicity axiom should be interpreted in that context. So, for example, the ``simplest" suggestion $h=0$ is vacuous, and as we shall see later, the constant $A$ in (\ref{h4}) must be chosen to be non-zero for nontrivial dynamics. Thus clearly one is searching for equations that have a minimum number of free parameters and yet  describe  interesting dynamical systems. This approach, of starting with the simplest nontrivial dynamics, allows for systematic extensions that are briefly discussed in the concluding section.

\end{itemize}

\section{Construction of the Hamiltonian}
Using the axioms one may construct an explicit form for $h(p,S)$. 
Rotational invariance and AHD imply that the building blocks of $h$ must be 
\begin{equation}
W_1,  \;\; g_{ij} U_1 (\partial_i U_2) (\partial_j U_3) \;\; \mbox{and} \;\; g_{ij} W_2 \partial_i \partial_j W_3 \,, \label{bb}
\end{equation}
where the $U_l, W_l, l=1,2,3$ are real functions of $p,S$ only (and not also of their derivatives). 
Separability restricts $h$ to be linear in $g_{ij}$: 
\begin{equation}
h = W_1 + g_{ij} \left( U_{1n} (\partial_i U_{2n}) (\partial_j U_{3n}) + W_{2n} \partial_i \partial_j W_{3n} \right) \,  \label{ht}
\end{equation}
where a possible additional index $n$ is summed over. 

Consider $W_1(p,S)$. Scale-invariance implies it cannot depend on $p$, so $W_1=W_1(S)$. But now separability requires $W_1 \propto S$. However in general the final result violates positivity of $H$ and hence one concludes that $W_1 =0$.

The terms involving $W_2$ and $W_3$ cannot generically lead to a positive definite density $h$ and so by the universality axiom we must set those terms to zero. Similarly, positivity of $h$ in the general case (universality) implies $U_{2n}$ =$U_{3n}$ and $U_{1n} \ge 0$ so that $h$ becomes a sum of positive terms. (One can actually avoid imposing the conditions in this paragraph and proceed as in \cite{P1}, but the discussion is more concise this way).

Hence  
\begin{equation}
h = g_{ij} U_{1n} (\partial_i U_{2n}) ( \partial_j U_{2n}) \, .  \label{h2}
\end{equation}

Next, using the chain rule, 
\begin{equation}
\partial_i U = \partial_i p { \partial U \over \partial p} + {\partial_i S} { \partial U \over \partial S} \, , 
\end{equation}
one can extract explicitly the derivatives of $p, S$ in the above expression. Then using the separability axiom one deduces, as in \cite{P1}, that various combination of terms must be  just constants.

Thus 
\begin{widetext}
\begin{equation}
h= g_{ij} \left( A (\partial_i S) (\partial_j S) + B {(\partial_i \log p) (\partial_j \log p)}  + \\
 \sum_n b_n (\partial_i(\log p + a_n S)) ( \partial_j (\log p + a_n S))      \right) \,  \label{h3}
\end{equation}
\end{widetext}
with $a_n \neq 0$ and $A,B,b_n$ non-negative. The result has been written in a form that emphasizes positivity. If one expanded out the terms in the sum over $n$, then the non-cross terms can be combined with the $A$ and $B$ forms, giving new non-negative coefficients $\bar{A}$ and $\bar{B}$, and leaving cross-terms involving $(\partial_i \log p)( \partial_j S)$ with a net coefficient $C=2 \sum a_n b_n$. Although the equations of motion would only depend on $\bar{A},\bar{B},C$ one would  still need to impose the positivity constraint which makes $C$ depend on $\bar{A},\bar{B}$ through the $a_n,b_n$.

Thus the dynamics appears to depend on the unlimited number of free parameters $a_n,b_n$. The simplicity axiom encourages one to reduce this dependence, in the first instance, by choosing $b_n \equiv 0$, thereby truncating $h$ to  
\begin{equation}
h= g_{ij} \left( A (\partial_i S)( \partial_j S) + B{(\partial_i \log p)( \partial_j \log p)} \right) \, . \label{h4}
\end{equation}  
A consequence of this diagonal form for $h$ is another simplification, for it leads to the usual continuity equation (\ref{conts}) where the particle velocities are independent of any explicit dependence on the probability density $p$. 
 
Since $A$ must be non-zero for  nontrivial dynamics, it can be absorbed in a redefinition of the metric. The final result for $h$ therefore depends only on a single new universal parameter $B$ with dimensions of action-squared. If $B=0$ then one has classical {\it ensemble} dynamics generalising the usual Hamilton-Jacobi description of classical mechanics. For nonzero $B$, the resulting hamilton's equations are, in the conventional normalisation $A=1/2$,  
\begin{eqnarray}
\dot{S} + {g_{ij} \over 2} \partial_iS \partial_j S + V + B  g_{ij} \left(  {\partial_i p \partial_j p \over p^2}  - {2 \partial_i \partial_j p \over p} \right) &=& 0 \, ,  \label{hj} \\
\dot{p}  + g_{ij} \ \partial_i \left( p \ \partial_j S \right)  &=& 0 \, .  \label{conts} 
\end{eqnarray}
These real nonlinear equations can be combined and rewritten, via the inverse Madelung transformation, as  the standard Schrodinger equation when $B$ is identified with ${\hbar}^2 / 8 $,
\begin{equation}
i\hbar \dot{\psi} = \left[ - {\hbar^2 \over 2} g_{ij} \partial_i \partial_j + V \right] \psi \, . \label{schmulp}
\end{equation}

Thus the same set of axioms have allowed us to obtain two theories: Classical ensemble dynamics for $B=0$ in (\ref{h4}) and the  non-relativistic  (and linear) Schrodinger equation for $B > 0$. The linear quantum theory is thus seen to be a single parameter extension of the classical theory \footnote{It should be noted that if one of the $b_n$ in (\ref{h3}) were nonzero then counting the corresponding $a_n$ means that one would have at least a {\it two}-parameter extension of classical dynamics, or equivalently a {\it non}-linear Schrodinger equation.}.

\section{Conclusion}

A main result of this investigation is that one may derive and understand the structure of Schrodinger's equation using intuitive classical concepts and axioms. In particular its linearity is seen to be a consequence of the other assumptions. 
Unlike  \cite{HR1,P1} where the quantum action was obtained by extending a given classical action, here both the classical and quantum dynamics were constructed from a single set of axioms. 

Although the axioms used here appear similar to those used in \cite{P1}, there are a number of crucial differences that should be highlighted. While positivity was used in \cite{P1} primarily to give the (inverse) uncertainty measure a sensible interpretation, the axiom $[A5]$ adopted here has been motivated by the need for a Hamiltonian bounded from below. In \cite{P1} scale-invariance of  $h$ was demanded as a sufficient condition for universality, while here scale-invariance has been argued to be a {\it consequence} of the broader requirement of universality. While the AHD condition was imposed in \cite{P1}, here the broader simplicity axiom has been used, of which AHD is a natural special case. Finally, the axioms in \cite{P1} were imposed on the (inverse) uncertainty measure that was added to the classical Lagrangian while here the axioms were used to construct the ensemble Hamiltonian.

As is manifest in Sect.(3), the universality and simplicity axioms have been used in ways that go beyond the related but narrower homogeneity and AHD conditions that were used in Ref.\cite{P1}. The broader conditions have been adopted so as to accomplish the wider scope of the construction: the full ensemble Hamiltonian in this paper versus a piece of the Lagrangian in Ref.\cite{P1}. Nonetheless, the more general axioms used here have a natural physical interpretation,  as discussed above, perhaps even more so than those used in Ref.\cite{P1}.

Further insight into the results of Sect.(3) can be obtained by enquiring about the type of equations that would result if one abandoned one or more of the axioms. For example, allowing higher number of derivatives of $p$ enables terms like \cite{P1}
\begin{equation}
h_1(p) = g_{ij} \partial_i (\log p + \eta f(p)) \partial_j (\log p + \eta f(p)) 
\end{equation}
with $ f(p)= g_{kl} (\partial_k \log p) (\partial_l \log p)$ and $\eta$ a constant. This $h_1$ satisfies all the constraints except AHD. Dimensional analysis shows that one must introduce a new length scale associated with such nonlinearities. Requiring universality implies that the nonlinear terms are associated with a universal length scale, a natural candidate being the Planck length. Thus in this way one sees a possible link between gravity and nonlinear corrections to Schrodinger's equation \cite{P2,P3}.  Such generalisations, and their interpretation in terms of short-distance physics, are discussed at greater length in Ref.\cite{P6} which also lists related literature.

A challenge is to extend the axiomatic construction to include fermions. This might involve the use of additional mathematical structures like Grassmann variables to summarise internal degrees of freedom, and a further refinement of the axioms.

\section{Appendix A: Galilean Invariance}

The equations of motion that follow from (\ref{h3}), for $V=0$, are 
\begin{eqnarray}
\dot{S} + \bar{A} {g_{ij}} \partial_iS \partial_j S + \bar{B}  g_{ij} \left(  {\partial_i p \partial_j p \over p^2}  - {2 \partial_i \partial_j p \over p} \right)- C g_{ij} \partial_i \partial_j S &=& 0 \, ,  \label{hjx} \\
\dot{p}  + 2 \bar{A} g_{ij} \ \partial_i \left( p \ \partial_j S \right) + Cg_{ij} \partial_i \partial_j p   &=& 0 \, ,  \label{contsx} 
\end{eqnarray}
with $\bar{A},\bar{B},C$ as defined earlier. One may wonder if demanding invariance of these equations under Galilean boosts, which have not yet been imposed, yields any constraints on the undetermined constants. 

As the essential features are manifest already for a single particle in one space dimension, consider that uncluttered case first. The relevant equations are 
\begin{eqnarray}
\dot{S} + {\bar{A} \over m} (\partial S)^2 + { \bar{B}  \over m}  \left(  {(\partial p)^2  \over p^2}  - {2 \partial^2  p \over p} \right)- {C \over m} \partial^2 S &=& 0 \, ,  \label{hjy} \\
\dot{p}  + {2 \bar{A} \over m} \ \partial \left( p \ \partial S \right) + {C \over m} \partial^2 p   &=& 0 \, ,  \label{contsy} 
\end{eqnarray}
and one would like these equations to be form invariant under the transformation $t' =t, x'=x-ut$, with $u$ the boost velocity. Since $p(x,t) dx$ is the probability of finding the particle in the region around $x$, it is required to be invariant under coordinate transformations. But $dx =dx'$ and so one deduces that $p'(x',t') =p(x,t)$ where $p'$ is the probability density in the primed frame. 

The transformation of the coordinates induces an obvious tranformation of the derivatives: $\partial / \partial x' = \partial / \partial x$ and $\partial / \partial t'= \partial / \partial t + u \partial / \partial x$. Start with the continuity equation in the primed frame and require it to have the same form as (\ref{contsy}), that is, with the unprimed quantities replaced by primed quantities. Then subtracting the primed equation from the unprimed equation and solving for  $S'$ gives the transformation rule,

\begin{equation}
S'(x',t') = S(x,t) - {mux \over 2 \bar{A}} + {m \over 2 \bar{A}} f(t) \int {dx \over p} + g(t) \, , \label{sprime}
\end{equation}      
where $f,g$ are functions of $t$ to be fixed next. 

Now compare the primed version of (\ref{hjy}) with the unprimed version and use (\ref{sprime}) to conclude   
\begin{eqnarray}
f(t) &=& 0 \, , \label{ff} \\
g(t) &=& {m u^2 \over 4 \bar{A}} t + \phi  \, , \label{gg}
\end{eqnarray}
where $\phi$ is a constant. Thus the non-trivial transformation of $S$ under Galilean boosts has been determined: notice that the result is independent of $\bar{B}$ and $C$! The independence from $\bar{B}$ is due to the fact that $\bar{B}$ multiplies  terms in the equations of motion which only depend on $p$ and its spatial derivatives and those structures are invariant by themselves  under boosts. The independence from $C$ can be understood as follows: The $C=0$ equations are invariant under (\ref{sprime}, \ref{ff}, \ref{gg}) which depend explicitly on $x$ only linearly. Thus the $C$ term in the equations of motion, which involves two spatial derivatives must be invariant under the same transformation.

Therefore one deduces that the transformation of the variables $p,S$ under Galilean transformations is the same in classical ensemble dynamics ($\bar{B} =0, C=0$), in linear quantum theory ($\bar{B} \neq 0, C=0$), and in the nonlinear theory with $C \neq 0$. The transformations do depend on the constant $\bar{A}$ but since that is just a normalisation factor it can be removed by redefining the metric (mass). 

Recalling the Madelung change of variables, it is not surprsingly that the transformation of $S$ found above is precisely the transformation of the phase of the Schrodinger wave function under Galilean boosts discussed, for example, in \cite{bal}. It should also be noted that the explicit linear dependence of $S'$ on the product $u x$ is consistent with the interpretation of $\partial S/m $ as the velocity of the particle mentioned in Sect.(1): it transforms correctly under boosts.

For the multidimensional case (\ref{hjx},\ref{contsx}) the transformations of the coordinates and derivatives under Galilean boosts are
\begin{eqnarray}
t' &=& t, \\
x_{i}^{'} &=& x_i -u_i t,\\
\partial_{i}^{'} &=& \partial_{i}, \\
\partial_{t'} &=& \partial_t + u_j \partial_j \, .
\end{eqnarray}
The metric $g_{ij}$ and probability density $p$ remain invariant. Then, as above, one deduces the transformation of $S$ (setting for convenience $\bar{A}=1/2$),
\begin{equation}
S'(x',t') = S(x,t) - \bar{g}_{ij} u_i x_j + { \bar{g}_{ij} u_i u_j \over 2} t + \phi  \,
\end{equation} 
where $\bar{g}$ is the inverse metric with diagonal coefficients $\bar{g}_{ii} = 1/g_{ii}$ and zero otherwise.  The constant parameter $\phi$ actually represents the global gauge invariance of the equations and corresponds to the conservation of probability.

\section{Appendix B: Gauge Inequivalence}
Setting $\bar{A} =1/2$ and $\bar{B} = \hbar^2 /8$, one may re-write the coupled equations (\ref{hjx},\ref{contsx}) in terms of the wavefunction $\psi = \sqrt{p} e^{iS/\hbar}$, 
\begin{equation}
i\hbar \dot{\psi} = H_s \psi  + F( \psi ) \psi \, , \label{nonsch}
\end{equation}
with $H_s$ the usual Schrodinger Hamiltonian and $F$ a nonlinear correction given by 
\begin{equation}
 F = C \left( {( \partial_i p) J_i \over p^2 } - { \partial_i J_i \over p} \right).
\end{equation}   
The current 
\begin{eqnarray}
J_i &\equiv& g_{ij} p \partial_j S \\
&=& {\hbar \over 2i} g_{ij} \left( \psi^{*} \partial_j \psi - \psi \partial_j \psi^{*} \right) \, 
\end{eqnarray}
 is that which appears in the continuity equation 
\begin{equation}
\dot{p} + \partial_i J_i =0 \, .
\end{equation}
The equation (\ref{nonsch}) belongs to a class of Galilean invariant nonlinear Schrodinger equations obtained in Ref.\cite{dgn}. In terms of the structures $R_1, R_4$ that are defined in \cite{dgn}, 
\begin{equation}
F = -C(R_1 -R_4) \, . 
\end{equation}

One may ask if the nonlinear piece $F$  may be eliminated through some change of variables in the equation (\ref{nonsch}), leading to a physically equivalent linear equation. The results of \cite{dgn} however show that for real nonlinearities the only nonlinear Schrodinger equations that are equivalent, through a nonlinear gauge transformation that keeps $p(x,t)$ invariant, to the linear Schrodinger equation are those for which the nonlinearity is  proportional to the usual quantum potential $Q$. Since the $F$ term above is not proportional to  $Q$, the nonlinear Schrodinger equation (\ref{nonsch}) is not equivalent to a linear Schrodinger equation.

It must be emphasized that although the nonlinear equation with $C \neq 0$  belongs to the class considered in \cite{dgn}, the  
equation derived here has some positivity constraints on the coefficients that come from the positivity imposed on the ensemble Hamiltonian, as discussed in the text. Such constraints are absent for the nonlinear equations in \cite{dgn} which were constructed using a different approach.

\end{document}